# Bipartite Network Model for Inferring Hidden Ties in Crime Data

Haruna Isah, Daniel Neagu, Paul Trundle
Artificial Intelligence Research Group
Department of Computing, University of Bradford
Bradford, UK
(H.Isah, D.Neagu, P.R.Trundle)@bradford.ac.uk

*Abstract*— **Certain crimes are hardly committed by individuals but carefully organised by group of associates and affiliates loosely connected to each other with a single or small group of individuals coordinating the overall actions. A common starting point in understanding the structural organisation of criminal groups is to identify the criminals and their associates. Situations arise in many criminal datasets where there is no direct connection among the criminals. In this paper, we investigate ties and community structure in crime data in order to understand the operations of both traditional and cyber criminals, as well as to predict the existence of organised criminal networks. Our contributions are twofold: we propose a bipartite network model for inferring hidden ties between actors who initiated an illegal interaction and objects affected by the interaction, we then validate the method in two case studies on pharmaceutical crime and underground forum data using standard network algorithms for structural and community analysis. The vertex level metrics and community analysis results obtained indicate the significance of our work in understanding the operations and structure of organised criminal networks which were not immediately obvious in the data. Identifying these groups and mapping their relationship to one another is essential in making more effective disruption strategies in the future.**

## I. INTRODUCTION

The Internet and related technologies lend themselves perfectly to crime coordination across dispersed areas [1]. The least common denominator of organised crime is human relationships, social networking is inevitable among criminal groups responsible for the provision of illicit goods and services [2]. Despite the ongoing efforts by governments, law enforcement agencies, academic researchers, and the security sector, little is yet known about the preferred structures, longevity, and how trust is assured among criminal groups [1]. Available empirical data suggest that conventional and cyber criminals are more likely to be involved in loosely associated illicit networks rather than formal organisations [1] [3]. Organised criminal groups often involve multiple offenders connected through various relationships [4]. These relationship can be represented as a network where the nodes are the criminals while the edges are the criminal interactions. Social network analysis, defined as a theoretical and methodological paradigm for sophisticated examination of complex social structures in [2], has a long history of application to evidence mapping in both fraud and criminal conspiracy cases [5], it is useful in understanding the patterns of relationships among criminal groups and in identifying key members in the group [6]. Node centrality and network density measures in social network analysis are useful in identifying pivotal nodes and potential fraud hotspots, sub-structures, structural holes and clustering coefficient measures are used for network classification and path prediction [7]. Link analysis allows for mixing of different node and edge types in the same network and is useful in generating investigative leads and for uncovering missing information that may be hidden in a criminal network [4]. Groups, also called communities or clusters in a network, can be considered as fairly independent compartments with high concentrations of edges within groups of vertices and low concentrations between these groups in the network [8]. Group detection is a useful method for understanding the structure and organisation of criminals in a network.

Criminal intelligence process relies on the ability to obtain and use data. Three main sources of data identified in [9] include: open data such as newsletters, closed data in form of structured databases and classified data often collected through covert means. A common starting point in understanding criminal groups is to identify the criminal's associates i.e. identifying relationships between individuals and their roles in the criminal activities [9]. These relationship are usually obtained from email and phone communication logs [10], underground forums [2] [11] [12] [13], scraped using set of seeds or leaked data [6] [14], money trails [15] [16], crime records [17], by extracting and associating entities in the grey literature [5] [18] or by combinations of these sources.

Situations may arise in a criminal dataset where there is no direct connection among the criminals, such is the case in [18] which involves the extraction of organisational structure of covert network from textual data obtained from public news, the archive of Evolution, an online black market in [13] that recently disappeared comprising of list of underground vendors with their associated products, and the crime report in [19] comprising of list of rogue manufacturers with their associated products. In order to address the issue of lack of direct connection among criminals and the need to understand their organisational structure, we propose the following research questions: (i) can we infer relationship based on common attributes and other metadata among entities involved in crime but not directly connected? (ii) is there an individual or groups directing the overall operations in a criminal network? We address these problems by (i) modelling a bipartite network in order to infer relationships between actors and resources involved in crime and (ii) analysing nodes and community structures of the resulting network.

This work is organised as follows: Section II provides a concise review of related work; Section III describes the research methodology; Section IV provides two case studies for evaluating our model and the final section describes conclusions and future work.

## II. RELATED WORK

Criminal network data may contain variety of entities such as persons, organisations, locations, URL's, vehicles, weapons, properties, bank accounts, etc. Learning associations between these entities is a critical part of uncovering criminal activities and fighting crimes [4]. Criminal groups show various levels of organisational structure. This organisation according to [1], depends on whether: (i) their activity is purely aimed at online targets such as swarms consisting of ephemeral clusters of individuals with no leadership as in the case of the Anonymous or hubs which are organised with a clear command structure and a focal point of core criminals around which peripheral associates gather as in the case of LulzSec, (ii) their activity uses online tools to enable conventional crimes such as clustered hybrids which are articulated around a small group of individuals and focused around specific activities or methods as in the case of carding networks or extended hybrids which are less centralised consisting of many associates and subgroups, and (iii) they combine online and offline targets such as hierarchies or aggregates according to their degree of cohesion and organisation. Social network analysis as a tool for understanding organised criminal groups involve the detection of structural changes in social networks with node and group level measures. The node level metrics include: degree and centrality measures while the group level metrics include: density, cohesion, group stability, etc. Group detection tasks in criminal network analysis involve detection of meaningful clusters of criminal actors, their interaction, the interaction of their subgroups and cliques in criminal data. Increased work in network analysis of criminal groups is reported recently in [2] [4] [5] [6] [10] [11] [12] [17] [18] [20] [21].

Criminals utilise underground forums in form of chatrooms and private messaging services to exchange information on abusive tactics and engage in the sale of illegal goods and services [2] [11] [12] [13]. Anonymised carding forums' private messaging records were modelled as graphs in [2] with the aim of uncovering the underlying structural and behavioural properties of cybercriminals, measures investigated include: degree distribution, assortativity, rich club phenomenon, transitivity and small world phenomena, connectivity and cohesive subgroups. Another analysis of underground forums aimed at understanding the social dynamics of six underground forums and how they impact e-crime market efficiencies was carried out in [11]. A recent study of underground forum interaction in [12] which uses six different centrality measures to produce a visual representation of cybercrime forum reports that criminal groups are organised into two distinct communities that resemble (i) gangs, which are limited in size with one central leader who makes all the decisions for the group and (ii) mobs, with hundreds or thousands of members that share relatively equal centrality rankings divided into multiple sub-groups. Vulnerability in organised crime groups that can be exploited by law enforcement agencies include how group members earn trust among pears and the way they get their money or the e-currency they use [12]. Market basket analysis of products traded in Evolution, an online black market operating on Tor network that recently became extinct was carried out in [13]. Results obtained from these studies would allow authorities to better utilise their resources and devise more effective disruption strategies in the future.

In [4], a link analysis technique that employs shortest path algorithms and priority first search to identify the strongest associations between entities in a criminal network was proposed and evaluated using Phoenix Police Department crime reports. The network of the 19 hijackers surrounding the tragic events of September 11th, 2001 were mapped through public data in [5], the result obtained revealed that the hijackers did not work alone but had accomplices who were on the planes, yet they were the conduits for money, skills and knowledge needed to execute the operation. Meta matrix model of concepts extracted from public news was used in [18] to detect and analyse the structure of a covert network.

Analysis of the community structure of Nigerian scammers were carried out in [6] and [10]. The study in [6] shows that these scammers are organised into tightly and loosely connected groups while the findings in [10] revealed that only ten groups are responsible for about 50% of the scam attempts we receive. In [17], n-clique and k-core algorithms compared fairly well with other propriety criminal group detection models. An analysis of cybercriminal ecosystem on Twitter in [20] reports that while network criminal hubs are more inclined to follow criminal accounts, the criminal accounts tend to be socially connected, forming a small-world. Network analysis of South African arms deal and corruption carried out in [21] reveals that a single actor can have many separate relationships through which resources (information, money, influence, etc.) are shared with other actors. If this actor were to be removed, a large part of the network would be disconnected from the rest and significant resources would be unable to reach large parts of the network.

Procedures for implementing social network analysis for organised crime prevention were described in [6] [7] [9]. The central theme in [6] is focused on constructing large scale social graph from a smaller set of leaked data and linking of the leaked data (set of email addresses) to Facebook profiles to scrape large scale social graphs. The main focus in [7] is centred around fraud analytics and the laid down procedures include: (i) building the network, (ii) graph sampling to select set of flagged nodes, (iii) exploring, observing and measuring fundamental network metrics, and (iv) applying mitigation measures based on the inference from the measured metrics. The main focus in [9] is operational, the procedure include understanding client needs, obtaining and use of relevant data, data quality evaluation, data collation strategy, data integration and analysis, knowledge dissemination, and finally, re-evaluation.

Although our approach also utilises social network measures for understanding criminal network grouping, its novelty can be traced to the use of bipartite network model, a special type of graph representation where vertices are divided into two sets $A$ and $B$, and only connections between two vertices in different sets are allowed [22]. The bipartite network representation naturally suits criminal datasets that may lack direct connection between criminals and allows for inferring of hidden ties among each of the two sets $A$ and $B$ as we can see in the next sections.

### III. METHODOLOGY

The research tasks are categorised as follows: (i) criminal network extraction, (ii) network representation, (iii) measuring network metrics, and (iv) analysis of group dynamics.

#### A. Criminal Network Extraction

One of the main challenges of data extraction for network analysis lies in the choice of vertices, relationships, and attributes that can best answer the targeted research questions. Candidate choices for each of the three major network elements are: vertices (individuals, groups, organisations, bank accounts, products, URL's, resources, affiliates, Internet infrastructures), edges (friendship, ownership, distributor/advertiser and can be binary, weighted, directed, undirected, multipartite or multiplex relationship) and attributes (location, time etc.).

Network elements extraction is straight forward in structured data such as police records where vertices have been structured in tables and the edges are derived either as binary or weighted relationship between single or multimode vertices. In a semi or unstructured data such as text documents, network constitute the union of all statements per text document, vertices are concepts or ideational kernels represented by one or more words, while edges are the links between two or more concepts [18]. In the latter case, data collection is more of an approximation via text or natural language network analysis.

#### B. Network Representation: The Bipartite Network Model

Let $A = \{a_1, a_2, \dots, a_i\}$, represent the vertex sets of actors such as individuals, group of individuals or organisations capable of initiating an action over certain resources and $B = \{b_1, b_2, \dots, b_i\}$, represent the vertex sets of resources such as products, then an actor is uniquely connected to resources and no connection exist between actors and actors or resources and resources. The set of edges or relationships $E \subseteq A \times B$, are the initiated actions while the edge weights represent the total co-occurrences of similar instance of connections. The pairs $a_x, b_y$ denote an actor $a_x$, who is associated to a resources $b_y$. The sets $A$, $B$, and $E$ can be represented as a bipartite graph $G = (A \mid B, E)$, where $A$ and $B$ are called the partite sets of the graph vertices that are connected by an edge iff $(a_x, b_y) \in E$, with $1 \leq x \leq i$ and $1 \leq y \leq j$ where $i$ is the number of unique actors and $j$ is the number of unique resources in the network. The cardinality or the number of edges in the bipartite graph is represented by $n = |E|$. The pairs $a_x, b_y$ can also be represented as $b_y, a_x$ denoting a resources $b_y$ associated by an actor $a_x$, hence the actor-resources network can be represented by a weighted undirected bipartite network $M_{ixj}$. In order to infer ties among actors in the network, we transform the bipartite network $M_{ixj}$ to its unipartite components $A$ and $B$ or actor-actor and resource-resource network respectively. The unipartite networks are obtained in a process called bipartite projection described in [22]. The bipartite to unipartite transformation process is illustrated below. Fig.1. is an actor-resource bipartite network representation where the two sets of vertices are differentiated by red (resources) and green (actors) colours. The unipartite components are obtained by selecting one of the sets of vertices and linking two vertices from that set if they were connected to the same vertices of the other sets.

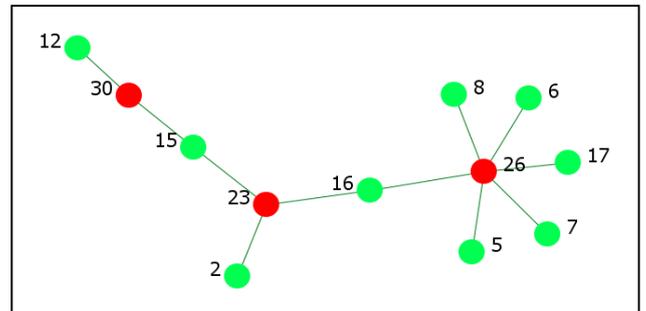

Fig.1. Actor-resource bipartite network

The $A$-projection of our actor-resource bipartite network $G = (A \mid B, E)$, shown in Fig.2. is the actor-actor network $G_A = (A, E_A)$ in which two vertices of $A$ are linked together if they have at least one neighbour in common.

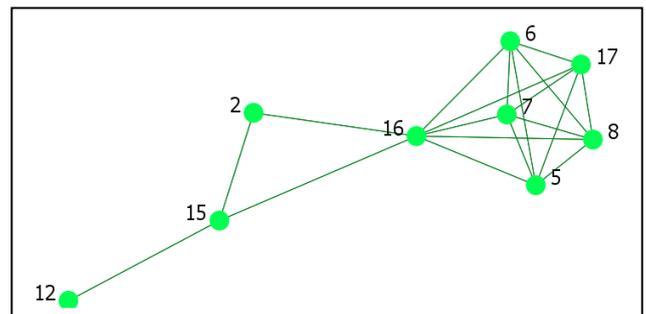

Fig.2. Actor-actor unipartite network

The $B$-projection of our actor-resource bipartite network $G = (A \mid B, E)$, shown in Fig.3. is the resources-resources network $G_B = (B, E_B)$ in which two vertices of $B$ are linked together if they have at least one neighbour in common.

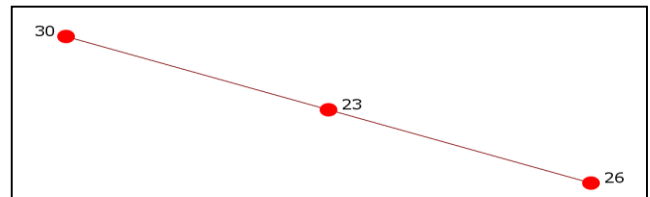

Fig.3. Resource-resource unipartite network

## C. Network Metrics

When trying to understand networks, we often want to identify important vertices, locate subgroups, or get a sense of how interconnected a network is compared to other networks. Vertex and edge specific measures include: degree, degree centrality, closeness centrality, betweenness centrality, eigenvalue centrality, PageRank and local clustering coefficient. Measures that can be used to describe the structure of the entire network include: density, degree distribution, connectivity and centralisation.

If the actor-actor network matrix $A$ is defined by:

$$a_{ij} = \begin{cases} 1 \text{ if an edge exists } from \text{ } vertex \text{ j } to \text{ } vertex \text{ i} \\ 0 \text{ otherwise} \end{cases},$$

then it follows from [6] that the degree of a vertex in the network defined as the number of edges connected to the vertex or the cardinality of the vertex neighborhood is given by:

$$d_i = \sum_j a_{ij} \tag{1}$$

the degree centrality is defined as:

$$D_i = \frac{d_i}{N-1} = \frac{\sum_j a_{ij}}{N-1} \tag{2}$$

where $d_i$ is the degree of the vertices and $N - 1$ is a normalization factor ($N$ is the number of vertices in the network) and $0 \leq D_i \leq 1$. Closeness centrality of a vertex, defined as the distance of a vertex from other vertices or the sum of shortest paths between a vertex and all other vertices in a network is given by:

$$C_i = (L_i)^{-1} = \frac{N-1}{\sum_j d_{ij}} \tag{3}$$

where $d_{ij}$ is the distance between vertices $i$ and $j$. $L_i$ is the normalized distance of a vertex from other vertices in a network. Betweenness centrality of a vertex, defined as the number of shortest paths in a network which passes through that vertex is given by:

$$B_i = \frac{\sum_{j<k} \frac{n_{jk}(i)}{n_{jk}}}{(N-1)(N-2)} \tag{4}$$

where $n_{jk}$ is the number of shortest paths between $j$ and $k$ and $n_{jk}(i)$ the number of such paths which pass through vertex $i$. $(N-1)(N-2)$ is the normalization factor. Eigenvector centrality of a vertex determines to what extent a vertex is connected to other well-connected vertices and is given by:

$$x_i = \frac{1}{\lambda} \sum_{j \in M(i)} x_j = \frac{1}{\lambda} \sum_j a_{ij} x_j \tag{5}$$

where $M(i)$ is the set of neighbors of $i$ and $\lambda$ is a constant. Clustering coefficient of a vertex is the probability that any two randomly chosen neighbours of that vertex in a network are connected themselves, hence a measure of the density of a 1.5-degree egocentric network. Density of a network, defined as a measure of how many edges are in a given set compared to the maximum possible number of edges in the network is:

$$density = 2 * \frac{|E|}{(|V| * (|V|-1))} \tag{6}$$

By counting how many vertices have each degree, a degree distribution is formed. Degree distribution deg(d) is defined as the fraction of vertices in a graph with degree d. Connectivity, also known as cohesion, is a count of the minimum number of vertices that would have to be removed before a network becomes disconnected. Centralisation uses the distribution of a centrality measure to understand the network as a whole. Once a network has been constructed and measurements have been calculated, the resulting dataset can be used for many applications.

## D. Community Structure

A useful way to understand a large network is to analyse some sections or subgraphs of the network referred to as egocentric networks. Subgraph allow us to identify common social roles and structures. Community detection in networks is a typical clustering problem [8] [23] and is aimed at identifying modules by using the information encoded in the network topology. Communities are assumed as groups of vertices that are similar to each other. This assumption allows for computing the similarity between each pair of the vertices with respect to some local or global reference property such that each vertex ends up in a cluster whose vertices are most similar to it, irrespective of whether they are connected by an edge or not [8]. It follows from [23], that in a network whose vertices can be assigned positions and embedded in an $n$ dimensional Euclidean space, the similarity or dissimilarity between the vertices can be computed using any norm $L_m$ such as: the Euclidean distance ($L_2$-norm), Manhattan distance ($L_1$-norm), or the $L_\infty$-norm. The concept of structural equivalence where similarity is inferred from the adjacency relationships between vertices is used for networks that cannot be embedded in space. Another important class of measures of vertex similarity is based on properties of random walks on graphs [8]. Such measure is the basis of the Walktrap algorithms in [23]. Another property of particular interest is whether or not all vertices in a group are connected to one another, when this happens, it is called a clique. A clique requires that all objects of a subgraph are connected to each other. A k-clique is a complete subset of size k of a graph [6].

## IV. EXPERIMENTAL WORK

We conducted two case studies: one on traditional counterfeiting crime (rogue manufacturer-manufacturer network), and the other on cybercrime (Darknet vendor-vendor network) data, in order to evaluate how we can infer ties among criminals using bipartite network modelling.

*A. Rogue Manufacturer-Manufacturer Network*

Pharmaceutical crime involves the manufacture, trade and distribution of fake, stolen or illicit medicines and medical devices, it also constitute the counterfeiting and falsification of medical products, their packaging and the associated documentation as well as theft, fraud, illicit diversion, smuggling, trafficking, illegal trade of medical products and the money laundering associated with it [3].

Pharmaceutical crime data may be composed of a variety of entities such as: people, organisations, brands, locations, storefronts, websites, bank accounts, and product delivery agencies [4]. These entities may form networks composed of: (i) thousands of storefronts in various locations (ii) affiliate websites run by associates', (iii) many botnet spamming partners who are paid to advertise illicit online pharmacy networks, (iv) covert systems for processing online orders, and (v) regular mail or courier services distributors, thereby making it difficult to track at the same time allowing the key actors to evade detection for long periods of time [3] [4]. Once these criminal groups are identified and their habits known, law enforcement authorities may begin to assess current trends in crime in order to forecast and hamper the development of perceived future criminal activities [9].

Network analysis of archived pharmaceutical crime data can be useful in modelling indirect relationships among important entities involved in pharmaceutical product counterfeiting. These entities can be criminals (manufacturers, advertisers, and distributors), the products they sell, banks that process their credit and debit card transactions or the delivery services used by these criminals. The method can also be used to reveal relationships between user accounts sending pharmaceutical spam and the spam URL's. The case study tasks include: (i) data extraction, (ii) network representation, (iii) vertices and network level analysis and (iv) group analysis.

*1) Rogue Data Extraction*

Using the year of sampling criteria, we extracted all the data from the Medicines Quality Database (MQDB), a public and freely accessible online tool that tracks medicines tested for quality in selected countries in Africa, Latin America and south-eastern Asia [19]. Currently, the database contains about 13,319 instances of medicines collected and tested from 12 countries. We extracted subset of the records with confirmed counterfeiting incidents. We then filtered duplicate data and removed all rows containing Missing, Unknown and N/A records in the Manufacturers column. We considered the following variables most relevant for the task at hand: Year, Manufacturer, Product, Country, Province, Dosage, Date of Sample Collection, and Test Result.

*2) Rogue Network Representation*

We constructed an undirected, weighted bipartite network, shown in Fig.4. between manufacturers (red vertices) with fake incidences and their associated products (green vertices) and called it rogue manufacturer-product network. The edge weight represent the co-occurrence frequency of manufacturer-product instances. The assumption we made here is that all manufacturers with atleast one product counterfeiting are rogues.

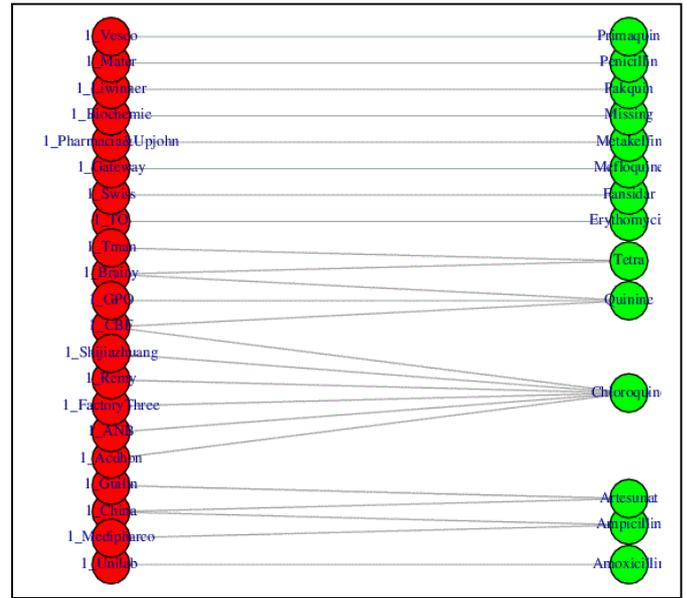

Fig.4. The rogue manufacturer-product network

The unipartite manufacturer-manufacturer network that resulted from projecting the network in Fig.4. is shown in Fig.5.

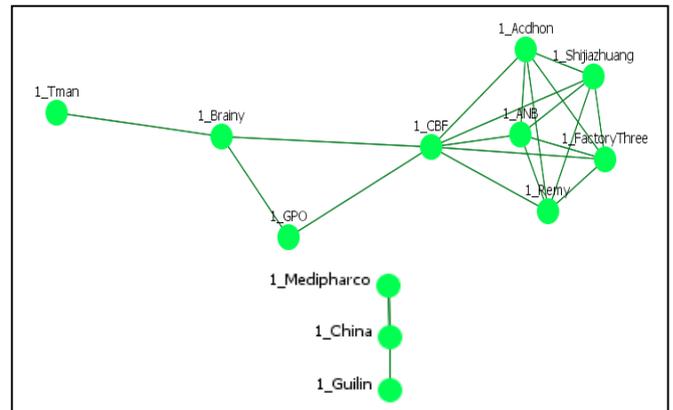

Fig.5. The rogue manufacturer-manufacturer network

*3) Rogue Manufacturer-Manufacturer Network Analysis*

We first report the aggregate metrics of the largest connected component of the network in Fig.5. These include: number of unique vertices = 9, number of unique edges = 19, geodesic distance (diameter) = 3, average geodesic distance = 1.4321, and network density = 0.5278. The vertex-specific network metrics for the larger component obtained by applying equation (1), equation (3), equation (4), and equation (5) for degree, betweenness, closeness and eigenvector centralities respectively are presented in TABLE I.

From the results in TABLE I, the important vertex in the rogue network is 1_CBF, it has the highest degree and centralities. A vertex with the most neighbours (degrees) can be said to be a key member with influence in its local neighborhood.

TABLE I. ROGUE NETWORK VERTEX-SPECIFIC METRICS

| Vertices | Degree | Betweenness | Closeness | Eigenvector |
|---|---|---|---|---|
| 1_FactoryThree | 5 | 0 | 0.083 | 0.149 |
| 1_ANB | 5 | 0 | 0.083 | 0.149 |
| 1_Shijiazhuang | 5 | 0 | 0.083 | 0.149 |
| 1_Remy | 5 | 0 | 0.083 | 0.149 |
| 1_CBF | 7 | 15 | 0.111 | 0.163 |
| 1_Acdhon | 5 | 0 | 0.083 | 0.149 |
| 1_GPO | 2 | 0 | 0.071 | 0.040 |
| 1_Brainy | 3 | 7 | 0.077 | 0.041 |
| 1_Tman | 1 | 0 | 0.050 | 0.008 |

*4) Rogue Manufacturer-Manufacturer Network Group Analysis*

We extracted 1.5 degrees egocentric networks of each vertex and reported subgraphs with more than three edges in Fig. 6. The star topology of egocentric network of the vertex 1_CBF indicates its relative importance as a switch or hub in the rogue network.

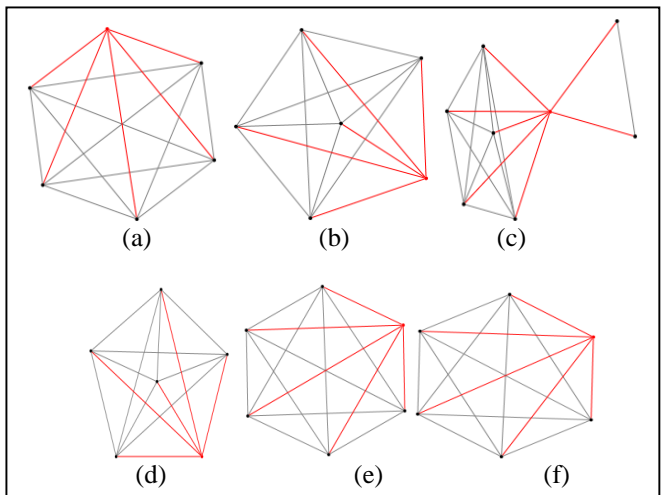

Fig.6. Subgraphs of the rogue network for the following vertices: (a) 1_Acdhon (b) 1_ANB (c) 1_CBF (d) 1_FactoryThree (e) 1_Remy (f) 1_Shijiazhuang

We applied four different community detection algorithms: Girvan Newman, Clauset Newman Moore, Wakita Tsurumi and Walktrap described in [8] in order to study the natural clusters in the rogue network. TABLE II is the summary of the community detection results for communities with minimum of three vertices. The result for the Walktrap method is presented in Fig.7.

When working with massive crime data with location and time attributes, these grouping might signal an element of organisation among the criminals. These naturally occurring clusters are based on patterns of social ties rather than formal group memberships. Vertices with a central position in their clusters, i. e. sharing a large number of edges with the other group partners, may have an important function of control and stability within the group while vertices lying at the boundaries between modules may play an important role of mediation and lead the relationships and exchanges between different communities.

TABLE II. ROGUE NETWORK COMMUNITIES

| Vertices in each cluster | Algorithms | | |
|---|---|---|---|
| | Girvan Newman | Clauset Newman Moore | Wakita Tsurumi |
| Cluster 1 | 1_FactoryThree 1_ANB 1_Shijiazhuang 1_Remy 1_CBF 1_Acdhon | 1_FactoryThree 1_ANB 1_Shijiazhuang 1_Remy  1_Acdhon | 1_FactoryThree 1_ANB 1_Shijiazhuang 1_Remy  1_Acdhon |
| Cluster 2 | 1_GPO 1_Brainy 1_Tman | 1_GPO 1_Brainy 1_Tman 1_CBF | 1_GPO 1_Brainy 1_Tman 1_CBF |
| Cluster 3 | 1_China 1_Guilin 1_Medipharco | 1_China 1_Guilin 1_Medipharco | 1_China 1_Guilin 1_Medipharco |

It is interesting to note that most of these incidents were recorded in one country.

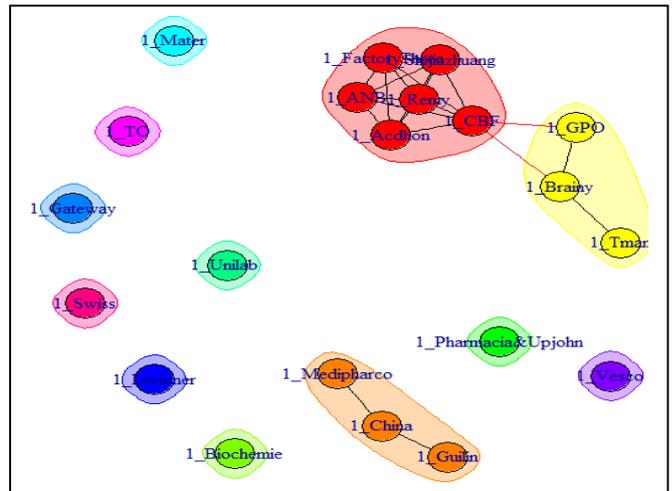

Fig.7. Rogue network communities using Walktrap algorithm

We further subject the rogue network to a more strict community detection methods so as to detect cliques. TABLE III is the clique community detection result.

TABLE III. NETWORK CLIQUE COMMUNITIES

| Communities | Rogue Cliques |
|---|---|
| 3-clique community | 21 |
| 4-clique community | 15 |
| 5-clique community | 6 |
| 6-clique community | 1 |

There were a total of 43 clique communities starting from a 3-clique community in the rogue network. The size of the largest clique community is 6 with a single clique. Clique communities in the rogue manufacturer-manufacturer network may lead to the detection of organised criminal network.

*B. Darknet Vendor-Vendor Network*

The Internet is now a catalyst for illicit online pharmacies, marketing falsified medicines to the public. There are several online market places on the hidden part of the internet (Darknets) offering prescription medicines together with cannabis for sale to the public, these market places such as Silk Road, Agora and Evolution provide online access anonymity via anonymising software such as P2P or Tor and payment anonymity via crypto-currencies such as Bitcoin to criminals with shipments being sent across the world between source, transit and destination countries [24]. Large-scale abusive advertising is a profit-driven endeavor, abuse-advertised goods and services such as spam-advertised Viagra, search-advertised counterfeit software and malware-advertised fake anti-virus have been dominated by an affiliate business model comprised of independent advertisers acting as free agents acquiring traffic via spam or search, and in turn paid on a commission basis by their sponsors who handle the back end, customer service and payment processing [15] [16]. Counterfeit pharmaceutical affiliate business models such as GlavMed, SpamIt and RX-Promotion involve a range of sponsors providing drugstore storefronts, drug fulfillment, shipping, payment processing, customer service and independent advertisers or affiliates who are paid a commission for promoting the program using botnets to send spam email or manipulating search engine results [15].

TABLE IV.  VENDOR-PRODUCT LISTING ON EVOLUTION DARKNET DATA

| Vendor | Products |
|---|---|
| MrHolland | Cocaine, Cannabis, Stimulants, Hash |
| Packstation24 | Accounts, Benzos, IDs & Passports, SIM Cards, Fraud |
| Spinifex | Benzos, Cannabis, Cocaine, Stimulants, Prescription, Sildenafil Citrate |
| OzVendor | Software, Erotica, Dumps, E-Books, Fraud |
| OzzyDealsDirect | Cannabis, Seeds, MDMA, Weed |
| TatyThai | Accounts, Documents & Data, IDs & Passports, PayPal, CC & CVV |
| PEA_King | Mescaline, Stimulants, Meth, Psychedelics |
| PROAMFETAMINE | MDMA, Speed, Stimulants, Ecstasy, Pills |
| ParrotFish | Weight Loss, Stimulants, Prescription, Ecstasy |
| Pharmacy_U | Analgesics, Drugs |

*1) Darknet Data Extraction*

We were opportune to have access to the archive of Evolution, an online black market operating on Tor network few weeks after it disappeared on March 18, 2015 via [13]. The extracted data consists of what product each vendor sells on different dates and contains few top-level categories such as Drugs, Digital Goods, Fraud Related, etc. which are subdivided into product-specific pages and each page contains several listings by various vendors as illustrated in TABLE IV. We extracted all records for Analgesics category and obtained 5002 records with the variables "Vendor", "Products", and "Date".

*2) Darknet Data Representation*

We constructed an undirected, weighted bipartite network, shown between vendors and their associated products and called it Darknet vendor-product network. The edge weight represent the co-occurrence frequency of vendor-product instances. We then transformed the network to its unipartite components, the vendor-vendor network obtained consists of 102 vertices, 952 edges and 4 connected components as shown in Fig.7.

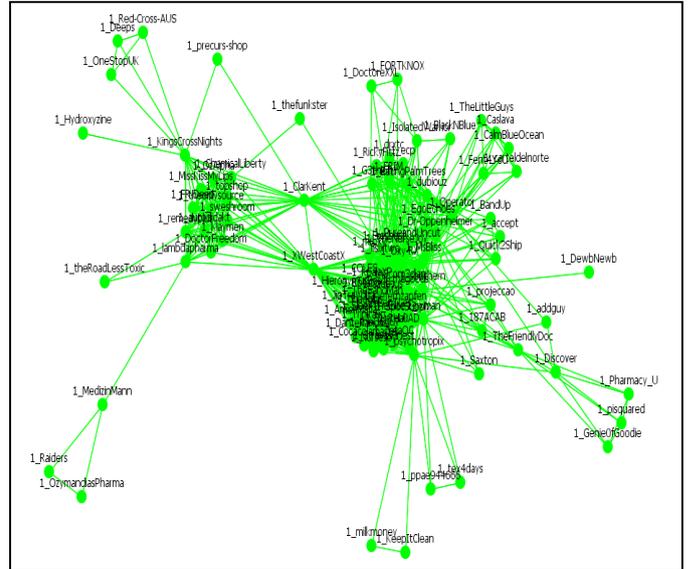

Fig.7. Darknet vendor-vendor network

*3) Darknet Vendor-Vendor Network Analysis*

We first report the aggregate metrics of the largest connected component of the network in Fig.7. These include: number of unique vertices = 95, number of unique edges = 947, geodesic distance (diameter) = 6, average geodesic distance = 2.1682, and network density = 0.2121. The important vendor in the Darknet network is 1_TheNurseJoy, it has the highest degree and centralities.

*4) Darknet Vendor-Vendor Network Group Analysis*

We obtained four clusters each of vertices 35, 26, 24, and 10 respectively when we applied Clauset Newman Moore community detection algorithms in [8]. There were however, 7 and 24 communities found with Wakita Tsurumi and Girvan Newman algorithms respectively. The size of the largest clique community is 28 with a single clique.

## V. CONCLUSION

Many types of crimes require a high degree of organisation and specialisation. Criminal networks show various levels of organisational structures, they may operate as: swarms, hubs, hierarchies, aggregates, or hybrids according to their activity, degree of cohesion and organisation. The organisation of crime may also occur in the darknet where individuals interact within online discussion forums and chat rooms. Identifying criminal groups within a network and mapping their relationship to one another can be essential to making intelligent strategic decisions. Social network analysis is being used to automatically identify criminal clusters that may not have been obvious in a crime dataset. These naturally occurring clusters are based on patterns of social ties rather than formal group memberships. Identifying these clusters and their boundaries allows for a classification of vertices according to their structural position in the groups. Vertices with a central position in their clusters may have an important function of control within the group while vertices lying at the boundaries between clusters may play an important role between different communities. Situations often arise in a criminal dataset where we lack direct connection among criminals. In this work, we model such data as a bipartite network in order to infer relationships between actors and resources based on their common attributes. We evaluated the model using two case studies and the results were very significant and can reveal some hidden ties among criminals that were not immediately obvious in the data.

We plan to undergo further evaluation of the model in a large scale case study and in collaboration with law enforcement agents. Weighted projection of the bipartite graph will be considered in the future.

ACKNOWLEDGMENT

Special thanks to the Commonwealth Scholarship Commission through the Association of Commonwealth Universities (ACU) for providing studentship to Haruna Isah.